\documentclass[showpacs,preprintnumbers,prd]{revtex4}
\usepackage{amsmath,amssymb}
\usepackage[dvips]{graphicx}

\newcommand{\W}{\mathcal{W}}
\newcommand{\tr}{\mathrm{tr}}
\newcommand{\Pcal}{\mathcal{P}}
\newcommand{\xt}{\boldsymbol{x}_\perp}
\newcommand{\yt}{\boldsymbol{y}_\perp}
\newcommand{\del}{\boldsymbol{\partial}_\perp}
\newcommand{\pt}{\boldsymbol{p}_\perp}

\begin{document}
\title{Randomness in infinitesimal extent
       in the McLerran-Venugopalan model}
\author{Kenji Fukushima}
\affiliation{Yukawa Institute for Theoretical Physics,
 Kyoto University, Kyoto 606-8502, Japan}
\begin{abstract}
 We study discrepancy between the analytical definition and the
 numerical implementation of the McLerran-Venugopalan (MV) model.
 The infinitesimal extent of a fast-moving nucleus should retain
 longitudinal randomness in the color source distribution even when
 the longitudinal extent approximates zero due to the Lorentz
 contraction, which is properly taken into account in the analytical
 treatment.  We point out that the longitudinal randomness is lost in
 numerical simulations because of lack of the path-ordering of the
 Wilson line along the longitudinal direction.  We quantitatively
 investigate how much the results with and without longitudinal
 randomness differ from each other.  We finally mention that the
 discrepancy could be absorbed in a choice of the model parameter in
 the physical unit, and nevertheless, it is important for a full
 theory approach.
\end{abstract}
\preprint{YITP-07-79}
\maketitle

%%%%%%%%%%   INTRODUCTION   %%%%%%%%%%

\section{Introduction}

  It is a tempting idea to approximate the wavefunction of an
interacting particle by mediating gauge fields surrounding the
particle.  Such a description of a fast-moving charged particle by
virtual photons is long known as the Weizs\"{a}ker-Williams
approximation.  If we apply this idea to the strong interaction, we
can approximately give the wavefunction of a heavy hadron by gluon
fields which extend from the local color distribution inside the
hadron.  The formalism of the Color Glass Condensate (CGC) can
accommodate the non-Abelian extension of the Weizs\"{a}ker-Williams
approximation~\cite{McLerran:1993ni,review}.

  The CGC formalism aims to embody the parton saturation picture
arising as a result of the small-x evolution at high
energies~\cite{Gribov:1984tu}.  The geometric scaling beautifully
indicates the existence of the saturation scale as a function of
Bjorken's x~\cite{Stasto:2000er}.  The CGC formalism has been forming
an important building-block of Quantum Chromodynamics (QCD) at high
energies.  With a Gaussian approximation for distribution of the
random color source and its dispersion given by the saturation
scale~\cite{Iancu:2002aq}, we reach the McLerran-Venugopalan (MV)
model~\cite{McLerran:1993ni}.

  So far, the MV model has been well examined analytically in the case
of a light projectile (e.g.\ a color dipole, a proton, etc) scattering
off a dense CGC
target~\cite{Kovchegov:1996ty,JalilianMarian:1996xn,Blaizot:2004wu,Blaizot:2004wv,Gelis:2005pt,Hatta:2006ci,Fukushima:2007dy}.
The scattering problem of two CGC
objects, i.e.\ high-energy nucleus-nucleus
collision~\cite{Kovner:1995ts,Gyulassy:1997vt,Kovchegov:1997ke,Fries:2006pv,Fukushima:2007ja}
is hard to solve analytically, however, and the possible analysis is
limited to the numerical
method~\cite{Krasnitz:1998ns,Krasnitz:2001qu,Lappi:2003bi,Lappi:2006fp,Lappi:2006hq}.
Only the initial fields right after the collision (i.e.\ initial
conditions) are analytically known in a simple
form~\cite{Kovner:1995ts,Krasnitz:1998ns,Fukushima:2006ax}.  The
initial chromo-electric and chromo-magnetic fields are thus calculable
by means of the proper Gaussian average over the color source
distribution~\cite{Fukushima:2007ja,Lappi:2006fp}.

  In this work we will address a problem that the numerical
formulation of the MV model assumes a crude approximation which leads
to a significant difference from the right answer.  We will clarify
the physical meaning to ascertain that the assumed approximation is
not acceptable.  The crucial point is that the formulation involves
two distinct limits in the longitudinal direction;  one is the
vanishing extent of a nucleus and the other is the vanishing
correlation length in the random color distribution.  We will then
quantify how much the numerical procedure turns out to underestimate
the expectation value of the initial fields as compared to the
analytical answer.

  One might wonder if the issue to be discussed here is academic and
only a technical detail, for phenomenological models are useful as long
as they can fit in with empirical data.  The existing numerical
simulations in the MV model are, in fact, fairly consistent with
experimental observations at RHIC (Relativistic Heavy-Ion Collider).
We would emphasize, however, that the MV model is not a fitting model
but a theoretical description based on QCD.\  Therefore, one should
not make little of correctness checking as a theory problem.

  Even though the numerical calculations underestimate the field
expectation value, one might suspect that its effect is negligible
from the fact that the numerical simulation agrees with RHIC
experimental data.  We shall see, however, that the approximation used
in the numerical simulation causes a factor over 10 in the initial
energy density.  Such a discrepancy is not visible in the former
numerical works in the MV model because the choice of the MV model
parameter $\mu$ can absorb it.  All the observables scale in accord
with $\mu$ in the saturation regime, and in the numerical simulation
usually, $\mu$ is determined so as to reproduce the particle
multiplicity.  As a result, for instance, twice larger $\mu$ leads to
16 times larger energy density which may compensate for the
discrepancy in the dimensionless coefficient.  This ``posteriori
reasoning'' would sound fine as a phenomenological approach.  The MV
model is a theory, however, as we emphasized above.  The essential
notion of the CGC idea is that a universal description for the hadron
wavefunction becomes possible once the saturation occurs.  Hence,
$\mu$ could be fixed independently by information obtained in the pQCD
calculation or in the deep inelastic scattering (DIS) through the
relation to the saturation scale $Q_s$ as a function of $\mu$.  [See a
recent nice work \cite{Lappi:2007ku} for careful discussions.]  It is,
of course, quite difficult to specify $\mu$ for RHIC in this way
because the relevant $Q_s$ at RHIC is not unique from the kinematics
unlike the DIS case.

  The reason we would stick to rigorousness also lies in the time
scale of early evolution after the heavy-ion collision.  A larger
$\mu$ may lift the underestimated energy and multiplicity up at the
cost of changing the time scale $\propto 1/\mu$.  Moreover, our
finding must have a significant effect on the ``Glasma'' instability
proposed in Ref.~\cite{Romatschke:2005pm}.  Because the correct
treatment in the longitudinal direction is important in our argument,
as we will see later, the instability with respect to the longitudinal
variable (rapidity) should naturally be affected.  Actually the
instability is found too weak in Ref.~\cite{Romatschke:2005pm}.  It is
likely that the numerical simulation in Ref.~\cite{Romatschke:2005pm}
underestimates the instability strength in the same way as for the
energy density.  If so, there might be a chance that the genuine
Glasma instability would occur faster to contribute more or less to
the early thermalization mechanism.

  One more point that motivates us to reveal the discrepancy stemming
from the numerical approximation is the infrared (IR) property of the
MV model.  The IR cutoff has a physical meaning, while the ultraviolet
(UV) cutoff should be zero in the end.  The dependence on the IR
cutoff strongly relies on the approximation.  This feature would bring
about uncertainty, as is discussed in Ref.~\cite{Lappi:2007ku}, in
addition to the overall dimensionless factor.

  In this work we will adopt the same prescription for the analytical
calculations as in the numerical simulation to regularize the UV and
IR singularities;  we use the lattice formulation with the spacing and
the system size given by $a$ and $La$, respectively.  This enables us
to make a direct comparison between the analytical and numerical
results.

%%%%%%%%%%   McLerran-Venugopalan MODEL   %%%%%%%%%%

\section{McLerran-Venugopalan Model}

  Here we will make a quick review on the MV model applied for the
relativistic heavy-ion collision (two-source problem).  If we have two
sources which represent particles moving fast in the positive and
negative $z$-directions (``right'' and ``left'' respectively), we can
write the corresponding current using the light-cone coordinates as
\begin{equation}
 J^\mu = \delta^{\mu+}\rho^{(1)}(\xt,x^-)
  + \delta^{\mu-}\rho^{(2)}(\xt,x^+) \,,
\label{eq:current}
\end{equation}
where we dropped $x^+$-dependence in the right-moving source and
$x^-$-dependence in the left-moving source as usual because of the
Lorentz time dilatation.  Then, converting the coordinates from the
light-cone variables $x^\pm$ to the proper-time $\tau$ and the
rapidity $\eta$ which are related by
$x^\pm=(\tau/\sqrt{2})e^{\pm\eta}$, we can express the initial fields
right after two nuclei collide
as~\cite{Kovner:1995ts,Fukushima:2006ax,Romatschke:2005pm}
\begin{equation}
 \begin{split}
 E^\eta &= ig\Bigl(\bigl[ \alpha^{(1)}_1, \alpha^{(2)}_1 \bigr]
           + \bigl[ \alpha^{(1)}_2, \alpha^{(2)}_2 \bigr]\Bigr) \,,\\
 B^\eta &= ig\Bigl(\bigl[ \alpha^{(1)}_1, \alpha^{(2)}_2 \bigr]
           + \bigl[ \alpha^{(2)}_1, \alpha^{(1)}_2 \bigr]\Bigr) \,,
 \end{split}
\label{eq:initial}
\end{equation}
and the transverse fields ($E^i$, $B^i$) are zero~\cite{Lappi:2006fp},
under the assumption that $\rho^{(1)}(\xt,x^-)$ and
$\rho^{(2)}(\xt,x^+)$ behave like $\delta(x^-)$ and $\delta(x^+)$
respectively near the light cone~\cite{Gyulassy:1997vt}.  This
assumption is the case because of the Lorentz contraction when two
nuclei travel at the speed of light.  In the above, the color
matrices, $\alpha^{(n)}_i$'s, are the gauge field configurations in
the radial gauge ($A_\tau\propto x^- A^++x^+ A^-=0$) created from each
color source $\rho^{(n)}$.  The solution must satisfy
$\partial_i \alpha^{(n)}_i = -\rho^{(n)}$ with a constraint that
$\alpha^{(n)}_i$ takes a pure gauge form so that its field strength is
vanishing.

  The covariant gauge is most convenient to get an explicit solution.
By rotating the solution obtained in the covariant gauge to the radial
gauge, we can write down the desirable solution
as~\cite{Kovchegov:1996ty};
\begin{equation}
 \alpha^{(1)}_i(\xt,x^-)
  = -\frac{1}{ig} V(\xt,x^-)\partial_i V^\dagger(\xt,x^-) \,,\quad
 \alpha^{(2)}_i(\xt,x^+)
  = -\frac{1}{ig} W(\xt,x^+)\partial_i W^\dagger(\xt,x^+) \,,
\label{eq:alpha}
\end{equation}
where the gauge rotation matrices are
\begin{equation}
 \begin{split}
 V^\dagger(\xt,x^-) &= \Pcal_{x^-}\exp\biggl[
  ig\int_{-\infty}^{x^-}dz^- \int d^2\yt\,G_0(\xt\!-\!\yt)\,
  \tilde{\rho}^{(1)}(\yt,z^-) \biggr] \,, \\
 W^\dagger(\xt,x^+) &= \Pcal_{x^+}\exp\biggr[
  ig\int_{-\infty}^{x^+}dz^+ \int d^2\yt\,G_0(\xt\!-\!\yt)\,
  \tilde{\rho}^{(2)}(\yt,z^+) \biggr] \,.
 \end{split}
\label{eq:gauge_rot}
\end{equation}
Here $\Pcal_{x^\pm}$ denotes the path-ordering with respect to
$x^\pm$.  We remark that $\tilde{\rho}$ in Eq.~(\ref{eq:gauge_rot}) is
the color source in the covariant gauge that is not identical to
original $\rho$ in the radial gauge.  Since the Gaussian weight is a
gauge-invariant function of $\tr[\rho^2]=\tr[\tilde{\rho}^2]$ (see
Eq.~(\ref{eq:gauss})), we do not have to discriminate $\rho$ and
$\tilde{\rho}$ practically.  We will thus not use the tilde any more
but just write $\rho$ regardless of gauge.  The two-dimensional
massless propagator is defined by $\del^2 G_0(\xt)=-\delta^{(2)}(\xt)$
which is singular in both the IR and UV sectors.  Let $a$ and $L$ be
the lattice spacing and the number of the lattice sites, and then we
can express $G_0(\xt)$ in the lattice regularization as
\begin{equation}
 G_0(\xt) = \frac{1}{2L^2}\sum_{n_i=1-L/2}^{L/2}
  \frac{\exp[i(x_1 2\pi n_1 /La + x_2 2\pi n_2 /La)]}{2-\cos(2\pi n_1
  /L)-\cos(2\pi n_2 /L)} \,.
\label{eq:propagator}
\end{equation}
The summation is supposed to exclude a singular point $n_1=n_2=0$
because we impose the global color neutrality $\rho^{(n)}(\pt=0)=0$,
so that the singular point $n_1=n_2=0$ does not appear at all in
Eq.~(\ref{eq:gauge_rot}).  It is easy to check that
Eq.~(\ref{eq:propagator}) is reduced to the standard expression,
$G_0(\xt)\to\int\!d^2\pt\,e^{i\xt\cdot\pt}/[(2\pi)^2\pt^2]$,
in the continuum limit where $a\to0$ and $L\to\infty$ are taken.

  The later-time dynamics is uniquely determined by the equation of
motion with the initial fields given by Eq.~(\ref{eq:initial}).  Any
physical observables are therefore given in terms of $V$ and $W$ in
principle, which we shall denote as $\mathcal{O}[V,W]$ generally.  We
can compute the expectation value by taking the average;
\begin{equation}
 \bigl\langle \mathcal{O}[V,W] \bigr\rangle
  =\int[d\rho^{(1)}][d\rho^{(2)}]\,\W^{(1)}[\rho^{(1)}]
  \W^{(2)}[\rho^{(2)}]\,\mathcal{O}\bigl[V,W]\bigr] \,,
\end{equation}
with the Gaussian weight,
\begin{equation}
 \begin{split}
 \W^{(1)}[\rho^{(1)}] = \exp\biggl[-\int dx^- d^2 \xt\,
  \frac{\tr[\rho^{(1)}(\xt,x^-)^2]}{g^2[\mu^{(1)}(x^-)]^2} \biggr] \,,
  \\
 \W^{(2)}[\rho^{(2)}] = \exp\biggl[-\int dx^+ d^2 \xt\,
  \frac{\tr[\rho^{(2)}(\xt,x^+)^2]}{g^2[\mu^{(2)}(x^+)]^2} \biggr] \,.
 \end{split}
\label{eq:gauss}
\end{equation}
The normalization of the color trace is understood as
$\tr[t^m t^n]=\tfrac{1}{2}\delta^{mn}$, where $t^m$'s are the
SU($N_c$) algebra in the fundamental representation.  The scale
$\mu^{(1)}(x^-)$ and thus $\rho^{(1)}(\xt,x^-)$ (or $\mu^{(2)}(x^+)$
and thus $\rho^{(2)}(\xt,x^+)$) may have finite extent in the $x^-$
(or $x^+$ respectively) direction since the Lorentz $\gamma$-factor
is finite in fact and also because the small-x evolution by the JIMWLK
equation should spread longitudinally in the color distribution.

  If we are interested in evaluating the initial energy
density in the heavy-ion collision, $\mathcal{O}[V,W]$ should be
$\tr[(E^\eta)^2+(B^\eta)^2]$, that
is~\cite{Fukushima:2007ja,Lappi:2006hq},
\begin{equation}
 \epsilon_0 =\bigl\langle \tr\bigl[(E^\eta)^2+(B^\eta)^2\bigr]
  \bigr\rangle = 2N_c(N_c^2-1)\,
  \langle\alpha^{(1)}\alpha^{(1)}\rangle
  \langle\alpha^{(2)}\alpha^{(2)}\rangle \,.
\label{eq:energy}
\end{equation}
where we introduced a notation to indicate the diagonal component,
i.e.\
$\langle\alpha_i^a\alpha_j^b\rangle=\delta_{ij}\delta^{ab}\langle\alpha\alpha\rangle$
for $\alpha^{(1)}$ and $\alpha^{(2)}$ respectively.  We will make use
of this notation again when we show the numerical results later.

%%%%%%%%%%   QUESTION   %%%%%%%%%%

\section{Question -- Numerical Approximation}

  In the numerical implementation, for practical reasons, it is
difficult to compute the expectation value of the Wilson
lines~(\ref{eq:gauge_rot}) as they are.  In the first approximation
the longitudinal finiteness is only negligible as compared to the
transverse size and one can take the following limit;
\begin{equation}
 \rho^{(1)}(\xt,x^-)\to \bar{\rho}^{(1)}(\xt)\,\delta(x^-)\,,\quad
 \rho^{(2)}(\xt,x^+)\to \bar{\rho}^{(2)}(\xt)\,\delta(x^+)\,.
\label{eq:limit}
\end{equation}
Then one might anticipate that all physical quantities are given as a
function of the integrated scale,
\begin{equation}
 \bigl[\bar{\mu}^{(1)}\bigr]^2 = \int_{-\infty}^{\infty}\!dx^-
  \bigl[\mu^{(1)}(x^-)\bigr]^2 \,, \quad
 \bigl[\bar{\mu}^{(2)}\bigr]^2 = \int_{-\infty}^{\infty}\!dx^+
  \bigl[\mu^{(2)}(x^+)\bigr]^2 \,.
\label{eq:scale}
\end{equation}
As a matter of fact, we can verify this expectation in the analytical
evaluation associated with only one
source~\cite{Blaizot:2004wu,Blaizot:2004wv,Fukushima:2007dy},
that is, only $[\bar{\mu}^{(1)}]^2$ (or $[\bar{\mu^{(2)}}]^2$)
appears in the Gaussian average of a function of $V$ (or $W$) with the
weight $\W^{(1)}[\rho^{(1)}]$ (or $\W^{(2)}[\rho^{(2)}]$
respectively).

  In the popular numerical formulation the Wilson
lines~(\ref{eq:gauge_rot}) simplify approximately under the
limit~(\ref{eq:limit}) as
\begin{equation}
 \begin{split}
 V^\dagger(\xt,x^-) &\overset{?}{\to} \bar{V}^\dagger(\xt,x^-)
  = \exp\biggl[ig\int\!d^2\yt\,G_0(\xt\!-\!\yt)\,
  \bar{\rho}^{(1)}(\yt)\,\theta(x^-) \biggr] \,,\\
 W^\dagger(\xt,x^+) &\overset{?}{\to} \bar{W}^\dagger(\xt,x^+)
  = \exp\biggl[ig\int\!d^2\yt\,G_0(\xt\!-\!\yt)\,
  \bar{\rho}^{(2)}(\yt)\,\theta(x^+) \biggr] \,.
 \end{split}
\label{eq:num_approx}
\end{equation}
These would be, of course, exact if there were not for the
path-ordering $\Pcal_{x^\pm}$ or if in the Abelian gauge theory.  The
question we are addressing in the present paper is the validity of
this naive prescription~(\ref{eq:num_approx}) that is commonly
employed in the numerical simulation.

%%%%%%%%%%   MATHEMATICAL REPRESENTATION   %%%%%%%%%%

\section{Mathematical Representation}

  Let us focus only on the $V$-sector in what follows because the
initial energy density is factorized into the $V$-sector and the
$W$-sector as is manifest in Eq.~(\ref{eq:energy}).  Needless to say,
exactly the same argument should work for the $W$-sector as well.  For
notational simplicity, then, we shall omit the superscript (1) which
represents the right mover.

  The subtle point comes from the fact that another Dirac delta
function is involved implicitly in the formalism besides
Eq.~(\ref{eq:limit}).  Namely, the Gaussian weight~(\ref{eq:gauss})
leads to the following correlation function;
\begin{equation}
 \bigl\langle\rho_a(\xt,x^-)\,\rho_b(\yt,y^-)\bigr\rangle
  = g^2\bigl[\mu(x^-)\bigr]^2\,\delta^{ab}\,
  \delta(\xt\!-\!\yt)\,\delta(x^-\!-\!y^-) \,.
\label{eq:correlation}
\end{equation}
We need to deal with two $\delta(x^-)$'s properly in order to
formulate the problem in question in an appropriate way.  For this
goal it is convenient to introduce a regularization to the longitudinal
Dirac delta functions.  We modify the Wilson line accordingly as
\begin{equation}
 V_\epsilon^\dagger(\xt,x^-) = \Pcal_{x^-}\exp\biggl[ig
  \int_{-\infty}^{x^-}\!dz^-\int\!d^2\yt G_0(\xt\!-\!\yt)\,
  \rho_\epsilon(\xt,z^-) \biggr] \,,
\label{eq:reg_wilson}
\end{equation}
where the regularized color source must be a smooth function
satisfying
\begin{equation}
 \lim_{\epsilon\to0}\rho_\epsilon(\xt,x^-)
  = \bar{\rho}(\xt)\,\delta(x^-) \,.
\end{equation}
We shall introduce another regularization for the correlation function
in a way as
\begin{equation}
 \bigl\langle\rho_a(\xt,x^-)\,\rho_b(\yt,y^-)\bigr\rangle_\zeta
  = g^2\bigl[\mu(x^-)\bigr]^2\,\delta^{ab}\,
  \delta(\xt\!-\!\yt)\,\delta_\zeta(x^-\!-\!y^-) \,,
\label{eq:reg_correlation}
\end{equation}
such that the regularized delta function must satisfy
\begin{equation}
 \lim_{\zeta\to0}\delta_\zeta(z^-) = \delta(z^-) \,.
\end{equation}

  We are now ready to elaborate the question in a mathematically
sophisticated manner.  The replacement of Eq.~(\ref{eq:num_approx}) is
justified if $\epsilon\to0$ comes first before $\zeta\to0$.  In the
analytical calculation, on the other hand, the relevant limit is
$\zeta\to0$ followed by $\epsilon\to0$ later.  Thus, a mathematically
sensible description of the question~(\ref{eq:num_approx}) should be
\begin{equation}
 \lim_{\zeta\to0}\lim_{\epsilon\to0}
  \bigl\langle \mathcal{O}\bigl[ V_\epsilon\bigr]\bigr\rangle_\zeta
 \overset{?}{=}
 \lim_{\epsilon\to0}\lim_{\zeta\to0}
  \bigl\langle \mathcal{O}\bigl[ V_\epsilon\bigr]\bigr\rangle_\zeta \,.
\label{eq:question}
\end{equation}
Here, the left-hand side corresponds to the numerical
implementation~(\ref{eq:num_approx}) and the right-hand side
corresponds to the one that is commonly assumed in the analytical
works~\cite{Blaizot:2004wu,Blaizot:2004wv,Gelis:2005pt}.

\begin{figure}
 \includegraphics[width=25mm]{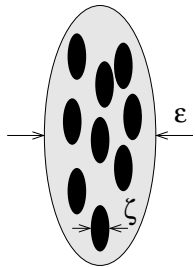}
 \caption{Schematic picture of a nucleus with two regulators
 $\epsilon$ and $\zeta$.  Intuitively $\epsilon$ signifies the
 longitudinal extent of the whole nucleus and $\zeta$ stands for the
 longitudinal correlation length inside the nucleus.  If randomness of
 the color source distribution is attributed to color confinement in
 each nucleon, $\zeta$ corresponds to the longitudinal extent of the
 nucleon.}
\label{fig:schematic}
\end{figure}

  We sketch the intuitive interpretation of $\epsilon$ and $\zeta$ in
Fig.~\ref{fig:schematic}.  Roughly speaking, $\epsilon$ is the
longitudinal extent of a fast-moving nucleus and $\zeta$ is the
correlation length of the color distribution inside the nucleus.  The
physical limit should keep $\epsilon>\zeta$ as is the case in the
right-hand side of the question~(\ref{eq:question}).  If $\epsilon$
goes to zero first, the longitudinal structure of randomness is lost.
Then only one infinitesimal ``sheet'' of the two-dimensional random
color distribution is left and the path-ordering becomes irrelevant
(see also Fig.~\ref{fig:slice}).  The numerical
prescription~(\ref{eq:num_approx}) hence implies such unphysical
ordering of two noncommutative limits.

%%%%%%%%%%   COMPARISON   %%%%%%%%%%

\section{Comparison}

  We will explicitly confirm that the order of two limits is
noncommutative indeed.  We will begin with the simplest example of the
tadpole expectation value.  Then we will proceed to the case of the
gauge fields which is directly related to the estimate for the initial
energy density by Eq.~(\ref{eq:energy}).  We note that the tadpole
gives the scattering amplitude between a quark in the color
fundamental representation and a CGC medium.  This scattering
amplitude is IR screened by long-ranged color interactions.  In
contrast to that, the gauge field correlation function involving four
Wilson lines contains a color-singlet component which is free from
screening.  However, in this case, the derivative acting onto the
Wilson lines is UV harmful leading to singular behavior.  It is
possible to take the limit of vanishing UV cutoff ($a\to0$) at finite
time ($\tau\neq0$) owing to non-linear evolution as has been closely
investigated numerically in Refs.~\cite{Lappi:2006fp,Lappi:2006hq} and
analytically in Ref.~\cite{Fukushima:2007ja}.  The initial energy
density in the heavy-ion collision at $\tau=0$ is, in this sense, an
ill-defined quantity in the $a\to0$ limit.  We will estimate it here,
nevertheless, to compare the analytical and numerical outputs.  It is
partly because, whether it is a physical quantity or not, we are
pursuing to exemplify a significant discrepancy anyhow.  We will do
this also partly because we know from Ref.~\cite{Fukushima:2007ja}
that the logarithmic singularity $\sim [\ln(La/a)]^2$ at $\tau=0$
translates into $\sim [\ln(La/\tau)]^2$ at $\tau\neq0$ with the
overall coefficient unchanged.  Thus, if we find a difference in the
overall factor of the initial energy density as we will do below, the
UV safe energy at $\tau\neq0$ should receive the same overall factor.
Accordingly the initial energy estimate here could serve as
correctness checking for UV safe observables indirectly.

%%%   Tadpole   %%%

\subsection{Tadpole}

  In the simplest case of the tadpole operator, i.e.\
$\mathcal{O}[V]=V^\dagger$, we can perform a quick comparison even
without resorting to the numerical method.  We already know the
analytical answer for the right-hand side of Eq.~(\ref{eq:question}).
That is given by~\cite{Blaizot:2004wv,Fukushima:2007dy}
\begin{equation}
 \lim_{\epsilon\to0}\lim_{\zeta\to0}\bigl\langle V_\epsilon^\dagger
  \bigr\rangle_\zeta = \exp\biggl[-g^4\bar{\mu}^2\frac{N_c^2-1}{4N_c}\,
  L(0,0)\biggr] \,,
\label{eq:rhs}
\end{equation}
where $L(0,0)$ is the notation in Ref.~\cite{Blaizot:2004wv} which is
defined as
\begin{equation}
 L(0,0) = \int d^2\xt G_0(\xt)G_0(\xt)
  = \frac{a^2}{4L^2}\sum_{n_i=1-L/2}^{L/2}\frac{1}
  {\bigl[2\!-\!\cos(2\pi n_1 /L)\!-\!\cos(2\pi n_2 /L)\bigr]^2}
  \simeq \frac{0.962 a^2}{2\pi}\biggl(\frac{L}{2\pi}\biggr)^2\,,
\end{equation}
in the lattice regularization.  Here, as in Eq.~(\ref{eq:propagator}),
the zero-mode ($n_1=n_2=0$) is to be removed by neutrality.  The
quadratic form approximates the sum quite well with a coefficient
$0.962$ that we find numerically.

  As for the left-hand side of the question~(\ref{eq:question}), we
have to perform the following Gaussian integral;
\begin{equation}
 \lim_{\zeta\to0}\lim_{\epsilon\to0}
  \bigl\langle V_\epsilon^\dagger\bigr\rangle_\zeta
  = \int[d\bar{\rho}]\, \exp\biggl[ig\int\!d^2\yt G_0(\xt\!-\!\yt)\,
  \bar{\rho}(\yt)\biggr]
  \exp\biggl[-\int d^2\xt \frac{\tr[\bar{\rho}(\xt)^2]}
  {g^2\bar{\mu}^2}\biggr] \,.
\label{eq:integral}
\end{equation}
It should be mentioned that the first exponential part is a matrix,
which makes the Gaussian integral hard to accomplish.  Although it is
a tough calculation for arbitrary SU($N_c$) group, the SU(2) case
($N_c=2$) is feasible immediately because the SU(2) exponential matrix
is easily manipulated.  After some calculations we find that the above
Gaussian integral results in
\begin{equation}
 \lim_{\zeta\to0}\lim_{\epsilon\to0}
  \bigl\langle V_\epsilon^\dagger\bigr\rangle_\zeta
  =\biggl(1-g^4\bar{\mu}^2\frac{1}{4}L(0,0)\biggr)
  \exp\biggl[-g^4\bar{\mu}^2\frac{1}{8}L(0,0)\biggr] \,,
\label{eq:lhs2}
\end{equation}
which obviously differs from Eq.~(\ref{eq:rhs}) with $N_c=2$
substituted;
\begin{equation}
 \lim_{\epsilon\to0}\lim_{\zeta\to0}
  \bigl\langle V_\epsilon^\dagger(\xt)\bigr\rangle_\zeta
  =\exp\biggl[-g^4\bar{\mu}^2\frac{3}{8}L(0,0)\biggr] \,.
\label{eq:rhs2}
\end{equation}
It is interesting to see from the above that Eq.~(\ref{eq:lhs2}) can be
a good approximation to Eq.~(\ref{eq:rhs2}) as long as
$g^2\bar{\mu} a L$ is small enough to allow for the Taylor expansion.

  It should be instructive to compute Eq.~(\ref{eq:integral})
numerically in the Monte-Carlo integration though we already have the
answer.  In order to compare to the existing numerical calculations in
literatures, we shall adopt the parameter choice same as used in
Ref.~\cite{Lappi:2006hq}.  That is,
\begin{equation}
 \frac{g^2}{4\pi} = \frac{1}{\pi}\,,\quad
 g^2\bar{\mu}a = 0.17\,,\quad L=700\,,
\end{equation}
was chosen in Ref.~\cite{Lappi:2006hq} and then the nucleus size is
given by
\begin{equation}
 R_A = \frac{a L}{\sqrt{\pi}}\,.
\label{eq:size}
\end{equation}
We change the value of $L$ to see the functional form.  Therefore, if
we increase $L$ while keeping $g^2\bar{\mu}a$ fixed at
$0.17$, the nucleus size or the infrared cutoff $R_A$ grows up in
proportion to $L$.

\begin{figure}
 \includegraphics[width=60mm]{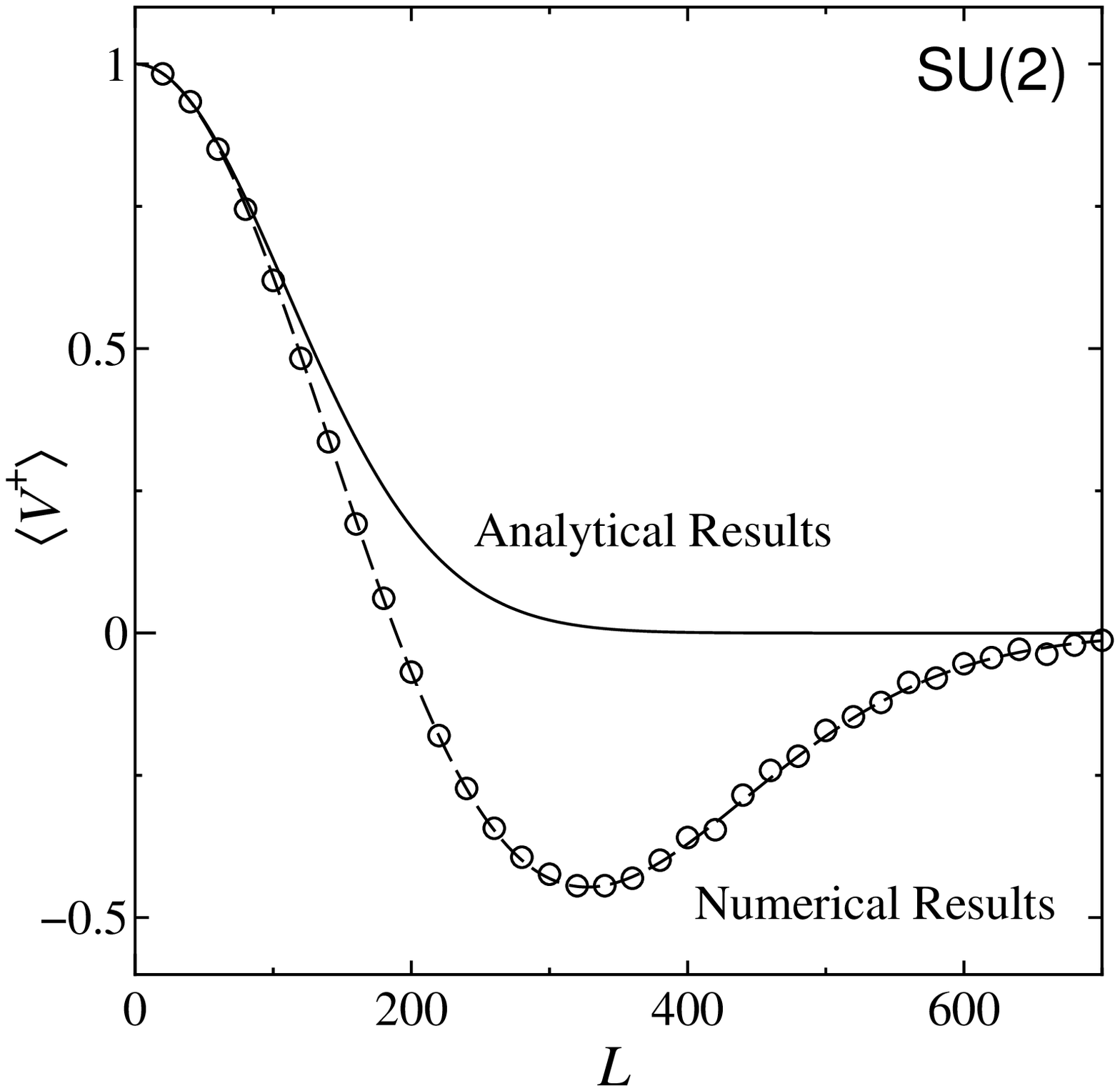}\hspace{10mm}
 \includegraphics[width=60mm]{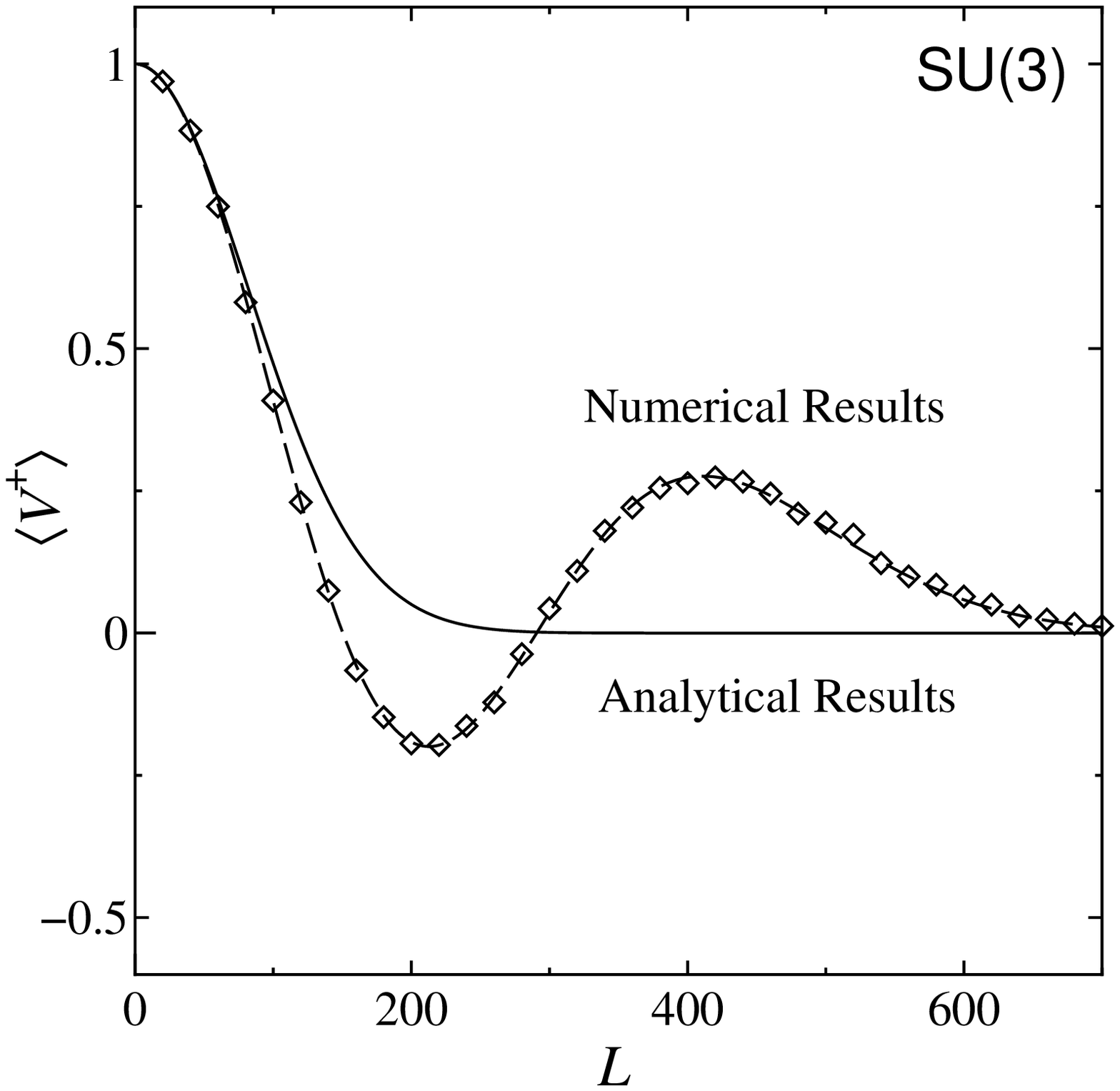}
 \caption{Tadpole expectation value as a function of $L$ with the
 parameter choice $g^2\bar{\mu}a=0.17$.  The left figure is for the
 SU(2) group and the right for the SU(3) group.  The solid curves
 represent the analytical results by Eq.~(\ref{eq:rhs}).  The open
 circles represent the numerical results by the Monte-Carlo
 integration.  The thin dashed curves show Eq.~(\ref{eq:lhs2}) in the
 SU(2) case and Eq.~(\ref{eq:lhs3}) in the SU(3) case, respectively.
 The SU(2) and SU(3) numerical data agree well with the thin dashed
 curves, which means that our Monte-Carlo integration works nicely.}
 \label{fig:tadpole}
\end{figure}

  We plot the analytical formula~(\ref{eq:rhs}) as a function of $L$
by the solid curve for the SU(2) case in the left of
Fig.~\ref{fig:tadpole} and the SU(3) case in the right of
Fig.~\ref{fig:tadpole}.  Since $g^2\bar{\mu}a$ is a constant,
what Fig.~\ref{fig:tadpole} means is the $R_A$-dependence of the
tadpole expectation value.  The open circles in Fig.~\ref{fig:tadpole}
represent the numerical results by means of the Monte-Carlo
integration.  We took 200 ensembles to calculate the expectation
value.  In the SU(2) case the numerical results agree well with the
expression~(\ref{eq:lhs2}).  Also, an analytical expression,
\begin{equation}
 \lim_{\zeta\to0}\lim_{\epsilon\to0}
  \bigl\langle V_\epsilon^\dagger\bigr\rangle_\zeta
  =\biggl(1-g^4\bar{\mu}^2\frac{1}{2}L(0,0)+g^8\bar{\mu}^4
  \frac{1}{24}L(0,0)^2\biggr)
  \exp\biggl[-g^4\bar{\mu}^2\frac{1}{6}L(0,0)\biggr] \,,
\label{eq:lhs3}
\end{equation}
can nicely fit the SU(3) results.  In any case, it is obvious that the
numerical results deviate from the analytical formula~(\ref{eq:rhs})
substantially.

%%%   Gauge Fields   %%%

\subsection{Gauge Fields}

  In view of the tadpole results, one might have thought that the
discrepancy is only minor.  The deviation may look small, however,
simply because the expectation value of color non-singlet operators is
exponentially suppressed by the system size $R_A$.  This fact becomes
manifest once we consider some other operators that contain a color
singlet component.

  Here, let us elucidate a more complicated situation than the
tadpole, that is, the expectation value of gauge fields
$\langle\alpha\alpha\rangle$ (see Eq.~(\ref{eq:energy}) for our
notation) which have a contribution from the color singlet.  In this
case the Gaussian integral is too tedious to accomplish, so we will
rely on the numerical Monte-Carlo integration only.  On the other
hand, the analytical calculation with the path-ordering is still
possible and it follows;
\begin{equation}
 \langle\alpha\alpha\rangle = \frac{1}{g^2}(g^2\bar{\mu})^2\,\sigma\,,
\label{eq:gauge_ana}
\end{equation}
where
\begin{equation}
 \sigma = \frac{1}{4L^2}\sum_{n_i=1-L/2}^{L/2}\frac{1}
  {2-\cos(2\pi n_1/L)-\cos(2\pi n_2/L)}
  \simeq \frac{1}{4\pi}\ln(1.36L) = \frac{1}{4}\ln(2.41R_A/a) \,,
\label{eq:sigma}
\end{equation}
which does not depend on $N_c$ at all and monotonically grows up with
increasing $L$.  We numerically find a constant $1.36$ in the
logarithm which approximates the sum.  This expression is, hence,
both IR and UV singular.  That is,
$\langle\alpha\alpha\rangle\to\infty$ whenever $L\to\infty$ (i.e.\
either $R_A\to\infty$ with $a$ fixed or $a\to0$ with $R_A$ fixed, see
Eq.~(\ref{eq:size})).  More importantly, the analytical
formulae~(\ref{eq:gauge_ana}) and (\ref{eq:sigma}) claim that the IR
behavior as $R_A\to\infty$ and the UV behavior as $a\to0$ are
completely identical.  This property is, however, no longer the case
in the numerical results with wrong approximation assumed, which could
be a possible explanation for IR stability found in
Ref.~\cite{Krasnitz:1998ns}.

\begin{figure}
 \includegraphics[width=60mm]{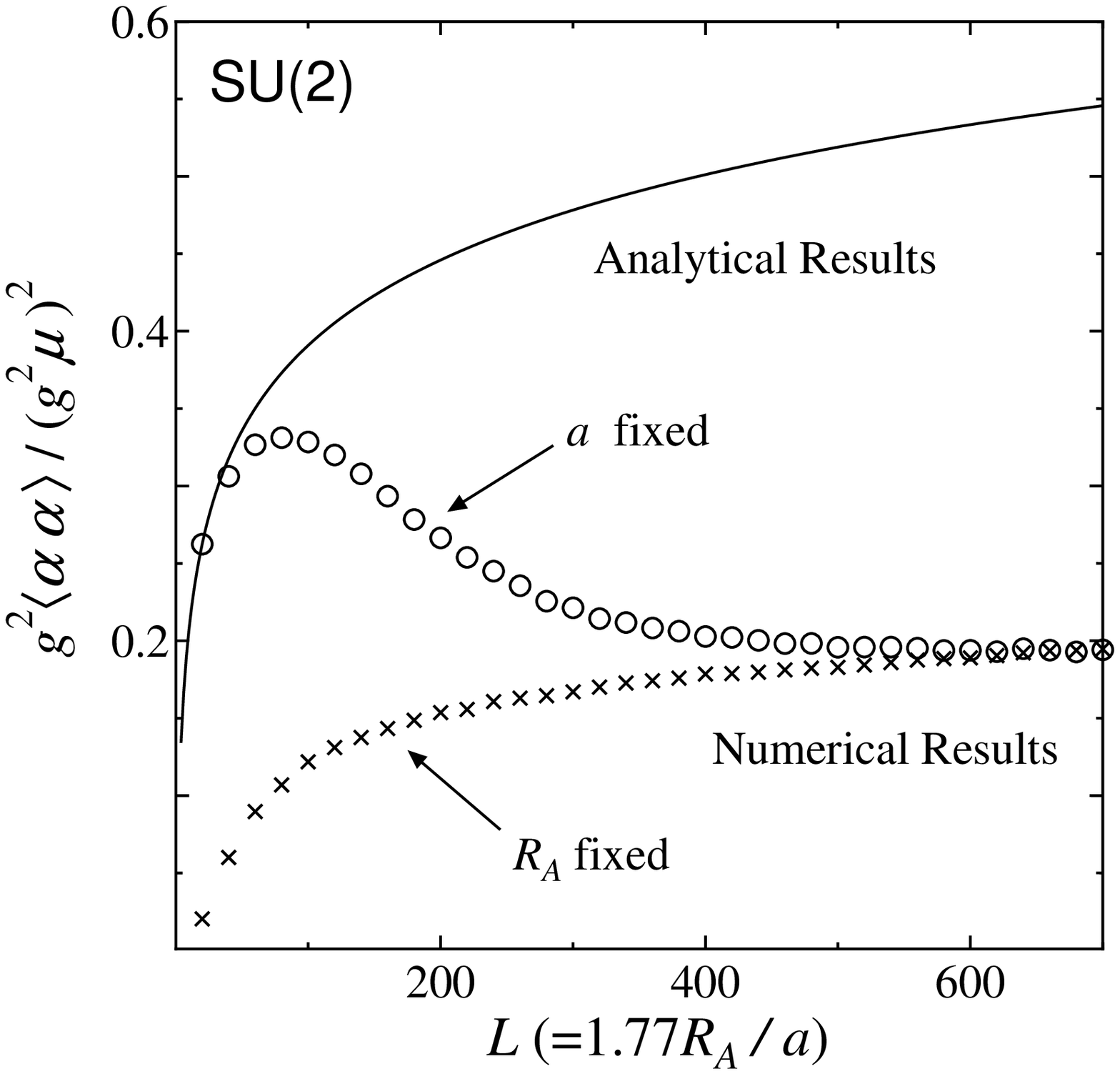}\hspace{10mm}
 \includegraphics[width=60mm]{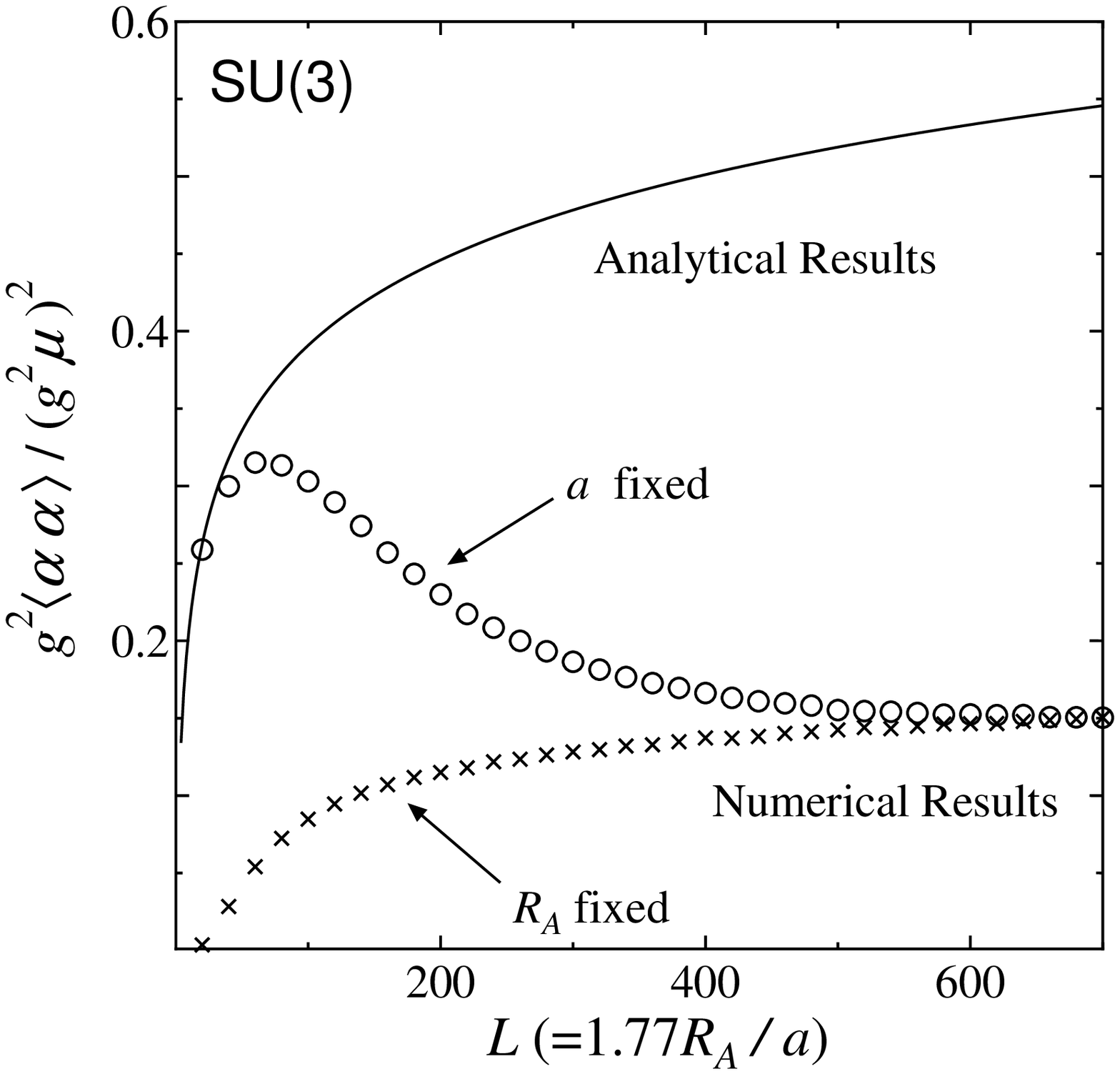}
 \caption{Gauge field expectation value as a function of $L$.  The
 analytical results are given by Eqs.~(\ref{eq:gauge_ana}) and
 (\ref{eq:sigma}).  The parameter $g^2\bar{\mu}a$ is fixed to be
 $0.17$ for the $a$-fixed results and $g^2\bar{\mu}a$ scales as
 $0.17\times700/L$ for the $R_A$-fixed results.  The left figure is
 for the SU(2) group and the right for the SU(3) group.}
 \label{fig:gauge}
\end{figure}

  Figure~\ref{fig:gauge} shows the results by the Monte-Carlo
integration.  The shape of
$g^2\langle\alpha\alpha\rangle/(g^2\bar{\mu})^2$ as a function of $L$
clearly depends on whether $R_A\propto L$ increases with $a$ fixed or
$a\propto 1/L$ decreases with $R_A$ fixed.  In the case that $a$ is
fixed, the calculation goes just in the same way as in the previous
subsection; we choose $g^2\bar{\mu}a=0.17$.  When we keep $R_A$ fixed,
we adjust the lattice spacing as $g^2\bar{\mu}a=0.17\times700/L$, so
that $g^2\bar{\mu}a$ becomes $0.17$ for $L=700$, which is completely
equivalent to the procedure in Ref.~\cite{Lappi:2006hq}.  Thus, the
$a$-fixed curve and $R_A$-fixed curve should meet at $L=700$, as is
certainly the case in Fig.~\ref{fig:gauge}.

  The numerical results hardly depend on $N_c$;  the left and right
figures of Fig.~\ref{fig:gauge} look almost the same, though the SU(3)
results are slightly smaller than the SU(2) ones.  The numerical
calculation leads to
$g^2\langle\alpha\alpha\rangle/(g^2\bar{\mu})^2=0.194$ at $L=700$ for
the SU(2) group, while the SU(3) group results in $0.150$ at $L=700$.
Differently from the tadpole case, the numerical data underestimate the
expectation value that is
$g^2\langle\alpha\alpha\rangle/(g^2\bar{\mu})^2=\sigma=0.546$ in the
analytical method.  In the next section, we will discuss how we may be
able to remedy this situation.

%%%%%%%%%%   Improvement   %%%%%%%%%%

\section{improvement}

\begin{figure}
\begin{minipage}[t]{.47\textwidth}
 \includegraphics[width=25mm]{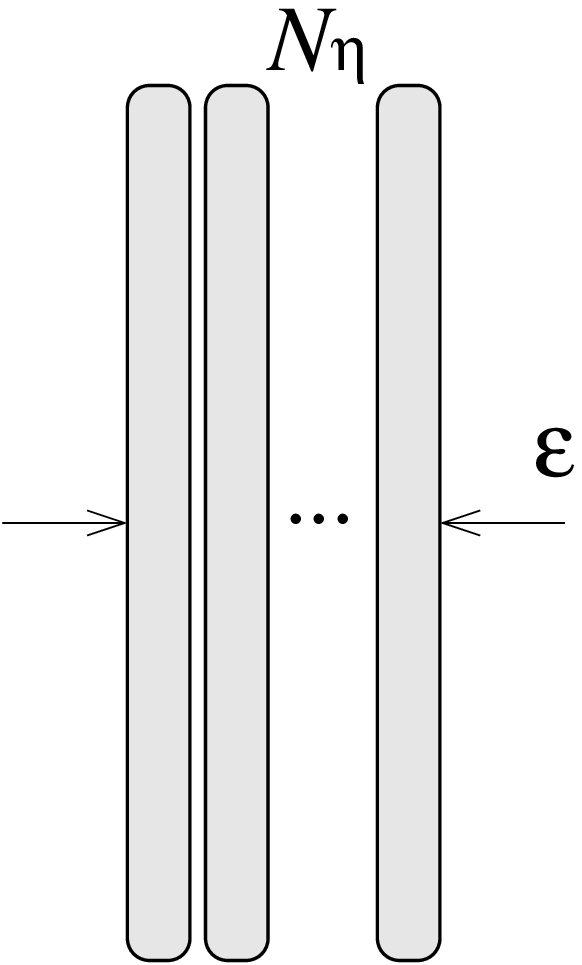}
 \caption{Schematic picture of how to improve the numerical results.
 Each blob represents the two-dimensional sheet without longitudinal
 randomness.  In the $N_\eta\to\infty$ limit we can recover the full
 randomness structure in longitudinal extent which is infinitesimal in
 the $\epsilon\to0$ limit.}
\label{fig:slice}
\end{minipage}
\hfill
\begin{minipage}[t]{.47\textwidth}
 \includegraphics[width=60mm]{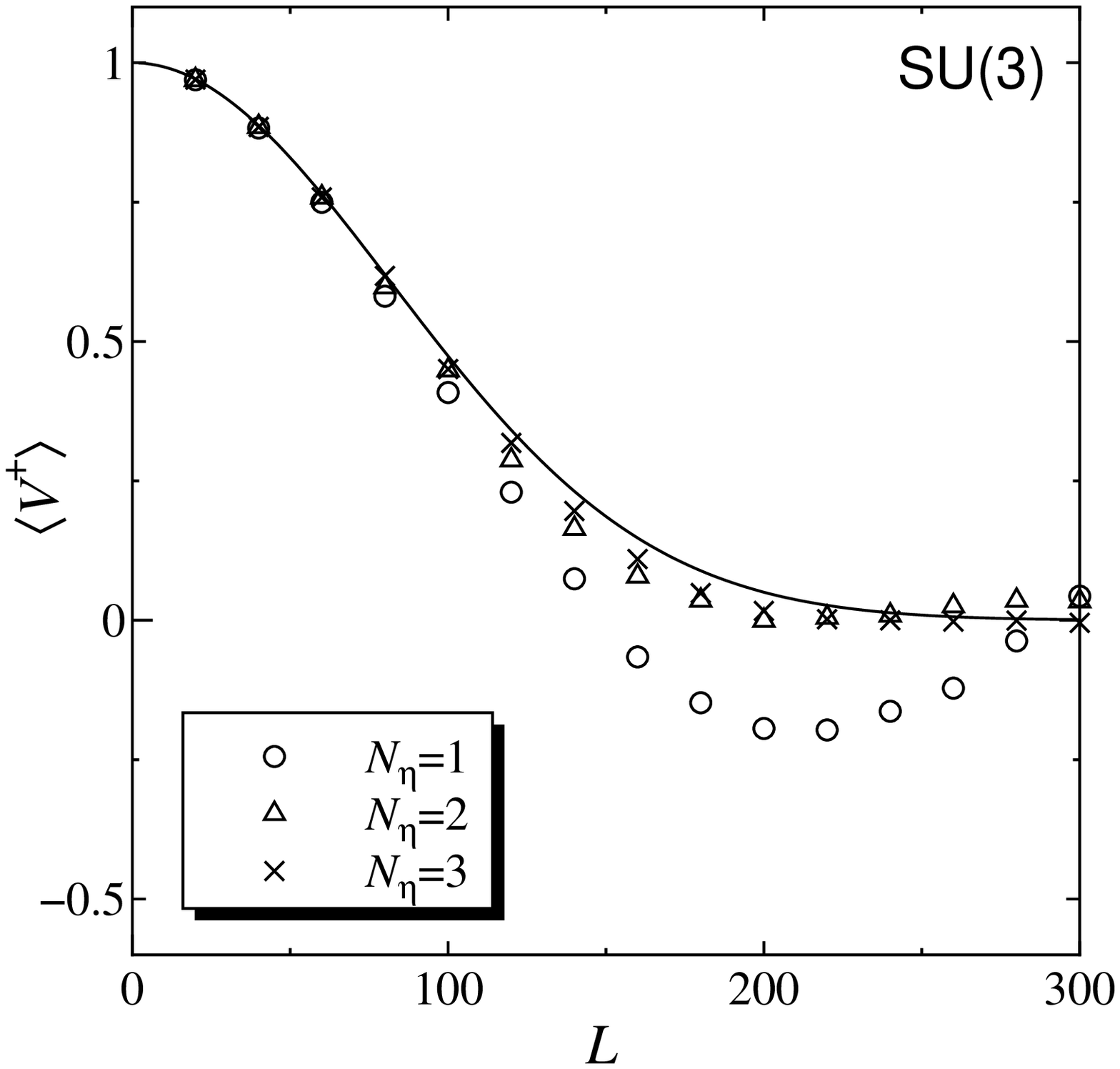}
 \caption{Tadpole expectation value for the SU(3) case at $N_\eta=1$
 (same as shown previously), $N_\eta=2$, and $N_\eta=3$.}
 \label{fig:tadpole_im}
\end{minipage}
\end{figure}

  We can improve the situation by inserting the infinitesimal sheet in
longitudinal extent in a way as sketched in Fig.~\ref{fig:slice}.  We
denote the number of sheets by $N_\eta$.  We note that $N_\eta$ is
\textit{not} the longitudinal coordinate as in
Ref.~\cite{Romatschke:2005pm}, but it is the number of slices within
infinitesimal extent.  Hence, all the numerical results we have seen
so far correspond to $N_\eta=1$.  We can recover the full
path-ordering in the $N_\eta\to\infty$ limit.

  For example, in the tadpole case, we shall modify the Gaussian
integral in Eq.~(\ref{eq:integral}) into the form of
\begin{equation}
 \bigl\langle V^\dagger\bigr\rangle_{N_\eta} = \prod_{n=1}^{N_\eta}
  \int\![d\bar{\rho}_n]\,\exp\biggl[ig\int\!d^2\yt
  G_0(\xt\!-\!\yt)\,\bar{\rho}_n(\yt)\biggr] \exp\biggl[
  -\int\!d^2\xt\frac{\tr[\bar{\rho}_n(\xt)^2]}
  {g^2(\bar{\mu}^2/N_\eta)}\biggr] \,,
\label{eq:improve}
\end{equation}
to retrieve the analytical results in the limit of $N_\eta\to\infty$.
We have to divide $\bar{\mu}^2$ by $N_\eta$ to make it consistent with
Eq.~(\ref{eq:scale}).  It is, however, impossible to take the
$N_\eta\to\infty$ limit in the practical procedure.  Instead, in this
section let us focus only on $N_\eta=2$, $N_\eta=3$, and $N_\eta=10$
to demonstrate the tendency of how the numerical outputs could move
from the structureless $N_\eta=1$ results toward the analytical
answer.  We did calculate in the SU(2) case as well as in the SU(3)
case, but we will present only the SU(3) results here, for the gauge
group makes no qualitative difference.

  The improvement works nicely for the tadpole expectation value as is
evident in Fig.~\ref{fig:tadpole_im}.  The results at $N_\eta=2$
almost reproduces the analytical curve.  So, the strategy to cure the
pathology might seem promising.

\begin{figure}
 \includegraphics[width=60mm]{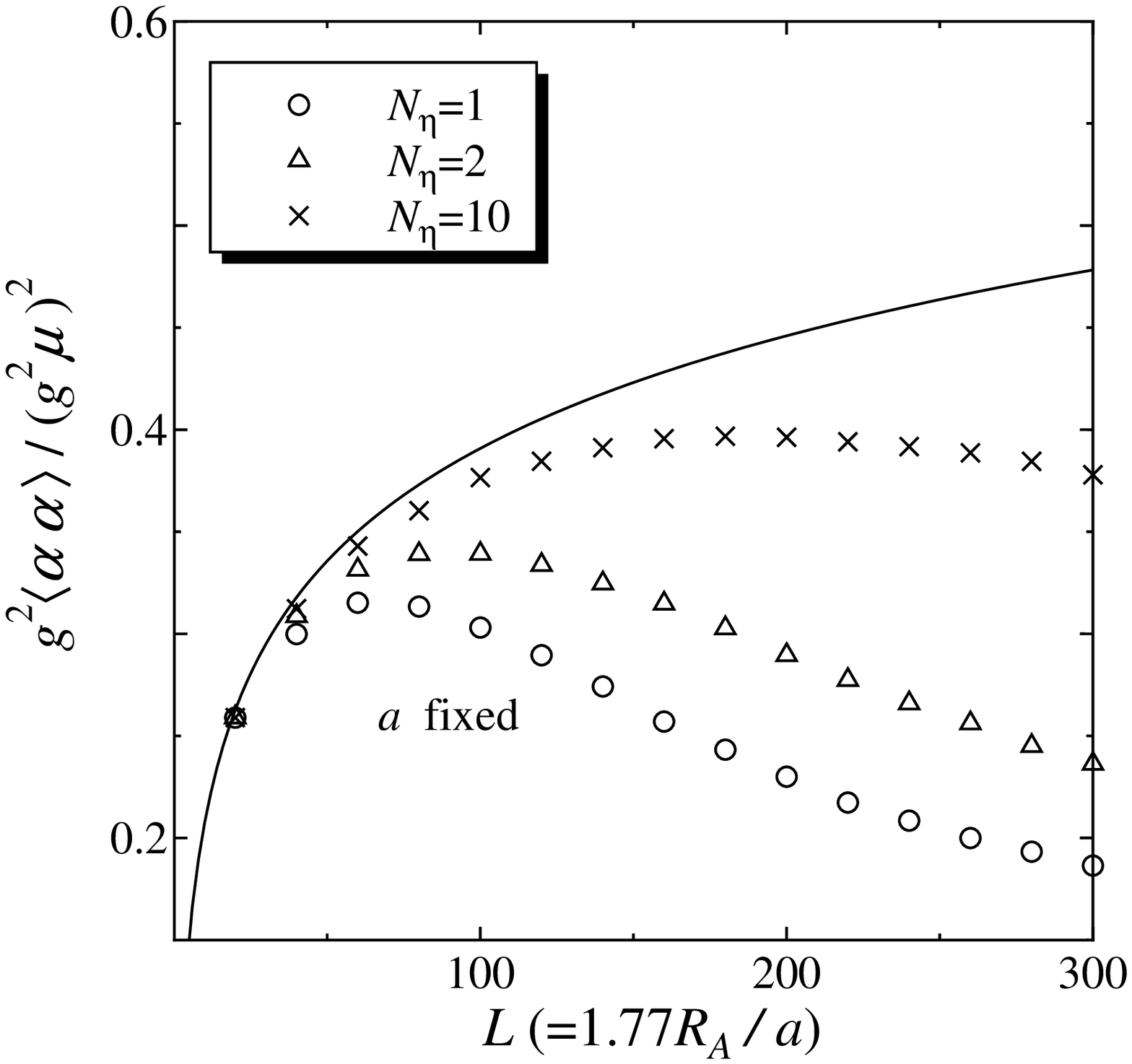}\hspace{10mm}
 \includegraphics[width=60mm]{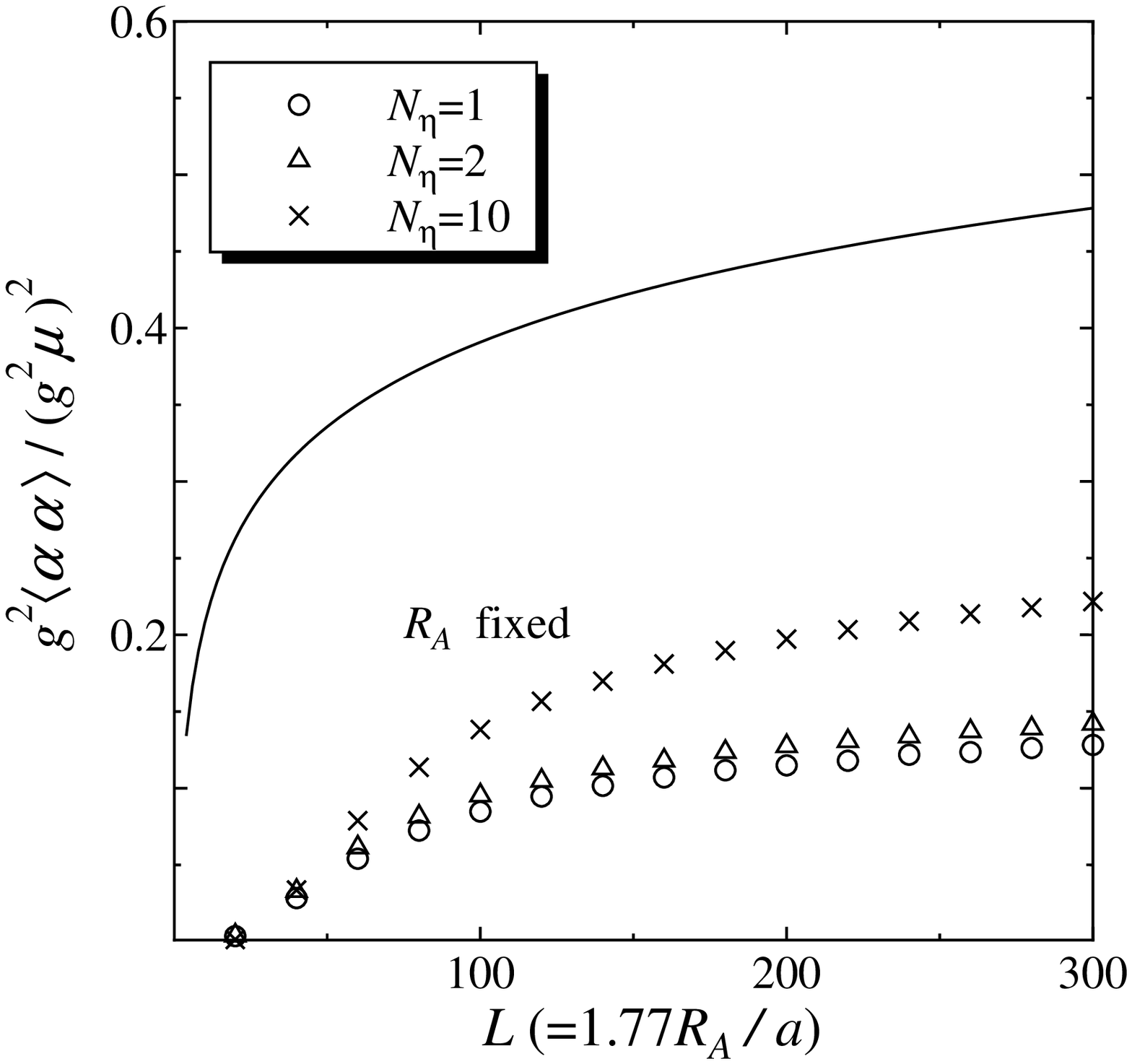}
 \caption{Gauge field expectation value for the SU(3) case at
 $N_\eta=1$ (same as shown previously), $N_\eta=2$, and $N_\eta=10$.
 The convergence to the analytical answer is much slower than the
 tadpole case especially for larger $L$.}
 \label{fig:gauge_im}
\end{figure}

  However, the gauge field expectation value scarcely benefits from
this improvement procedure.  We show the results in
Fig.~\ref{fig:gauge_im} for the $a$-fixed calculation (left) and the
$R_A$-fixed calculation (right).  The improvement seems rather better
for the IR behavior with $a$ fixed.  The UV behavior with $R_A$ fixed
barely moves with increasing $N_\eta$.  This is because the reference
point for the $R_A$-fixed calculation is chosen at $L=700$ where the
deviation between the analytical and numerical results is acute as
perceived from Fig.~\ref{fig:gauge}.  So, $N_\tau$ must be comparable
to $\sim700$ to achieve a nice deal of improvement for the $R_A$-fixed
results.  We would observe better convergence if $g^2\bar{\mu}aL$ is
smaller than $0.17\times 700$ that we chose here.

%%%%%%%%%%   DISCUSSIONS   %%%%%%%%%%

\section{Discussions -- From a ``Model'' to a ``Theory''}

  From the comparison between the analytical and numerical outputs, we
can learn an important lesson;  the numerical implementation like
Eq.~(\ref{eq:num_approx}) needs more and more caution as we approach
the continuum limit in the transverse plane.  This is a sort of
dilemma.  We should insert more and more sheets along the longitudinal
direction as in Fig.~\ref{fig:slice} when we want to make use of a
finer lattice or a larger volume in the simulation.  Of course, the
computation time increases as $N_\eta$ gets larger because the number
of the integration variables is proportional to $N_\eta$ as seen in
Eq.~(\ref{eq:improve}).

  We shall apply our results to the comparison of the initial energy
density between the analytical formula by Eq.~(\ref{eq:energy})
and what was reported in Ref.~\cite{Lappi:2006hq}.  As we have
discussed, the numerical formulation with the
approximation~(\ref{eq:num_approx}) underestimates the gauge field
expectation value by a factor $0.546/0.15=3.64$.  Therefore, the
initial energy density obtained in the numerical method is smaller
than the analytical estimate by a factor $3.64^2=13.2$.

\begin{figure}
 \includegraphics[width=70mm]{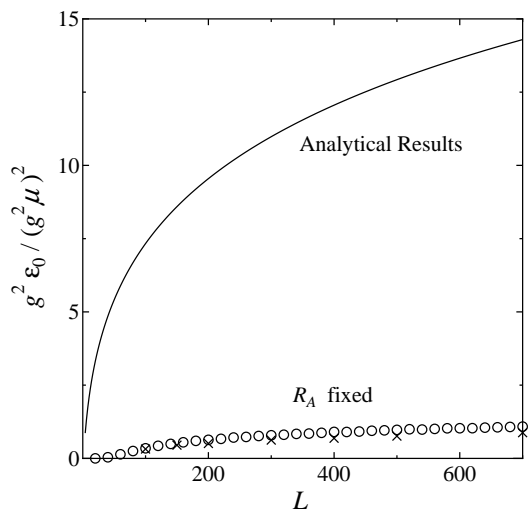}
 \caption{Initial energy density in the analytical calculation shown
 by the solid curve and the numerical estimate shown by the open
 circles.  The cross points represent the data presented in
 Ref.~\cite{Lappi:2006hq}.}
 \label{fig:energy}
\end{figure}

  We plot the initial energy density in Fig.~\ref{fig:energy}.  It
should be mentioned that, though the important message from
Ref.~\cite{Lappi:2006hq} is that the initial energy density at
$\tau=0$ is ill-defined and logarithmically divergent in the
$R_A\to\infty$ limit, our analytical calculation ends up with a finite
value with the IR and UV cutoffs (i.e.\ $a>0$ and $L<\infty$), which
is just the same as the lattice discretized results.

  It is notable that our numerical estimate is close to the data
calculated in the numerical simulation in Ref.~\cite{Lappi:2006hq}
which is overlaid on Fig.~\ref{fig:energy} by the cross points.  The
small discrepancy should be explained by difference between the naive
discretization in this work and the gauge-invariant lattice
formulation in terms of the link variables.

  One should not jump to a conclusion, however, that the initial energy
density becomes $13.2$ times larger than the estimate in the previous
numerical works.  To earn a right interpretation, we have to
understand how to specify the MV model parameter $\bar{\mu}$.  The
determination of $\bar{\mu}$ has been carefully explained in the
pioneering works~\cite{Krasnitz:1998ns,Lappi:2003bi} and their choice
($g=2$ and $g^2\bar{\mu}=2\;$GeV for RHIC) has become a standard.
Section~V~A in Ref.~\cite{Lappi:2003bi} is actually devoted to
elaborating that $\bar{\mu}=0.5\;$GeV can reproduce the total
multiplicity at RHIC and then the following energy density is in a
reasonable range.  This procedure would make the discrepancy between
the analytical and numerical results hidden behind the fitting;  not
only the energy density but also the multiplicity is accompanied by
the factor $3.64$.  [The multiplicity is an integration over the gauge
field expectation value divided by the particle dispersion relation.]
That means that the value of $\bar{\mu}$ would become
$\sqrt{3.64}=1.9$ times larger in order to fit the total multiplicity
without the overall factor $3.64$, leading to $3.64^2=13.2$ times
larger energy density.  Consequently, even though the dimensionless
coefficient has a substantial factor, the dimensional quantity
\textit{in the physical unit} remains unchanged.  In other words the
saturation model has only one dimensional scale $\bar{\mu}$ and all
the physical quantities are expected to scale with $\bar{\mu}$.  Once
$\bar{\mu}$ is fixed by means of one of the experimental data set,
other observables from the model calculation should be all consistent
with the whole data set.

  Then, one might be confused at what the point of our finding is in
this work.  Our aim is, as emphasized in Introduction, to establish a
correct \textit{theoretical} procedure for the MV model.  For that
purpose the phenomenological success is inadequate.  We must treat the
MV model scale $\bar{\mu}$ not as a fitting parameter but as an input
deduced theoretically from a given Bjorken's $x$.  This is, in
principle, possible with the BFKL equation from which the saturation
scale $Q_s$ is inferred as a function of $x$.  [The nonlinear
extension of the BFKL equation, namely the JIMWLK equation, is not
necessary for the determination of $Q_s(x)$ as concisely explained in
Ref.~\cite{Kharzeev:2004if}.]  The theoretical uncertainty by higher
order corrections is still large, however, and the
Golec-Biernat-W\"{u}sthof fit multiplied by the atomic number factor
$A^{1/3}$ is usually regarded as a more trustworthy form of $Q_s(x)$.
Then, within the framework of the MV model, the saturation scale can
be defined in terms of $\bar{\mu}$ as
\begin{equation}
 Q_s^2 \simeq \frac{N_c g^2\bar{\mu}^2}{4\pi}\ln(Q_s^2/\Lambda^2) \,,
\label{eq:satu_MV}
\end{equation}
where $\Lambda$ is the IR cutoff.  That is, $\Lambda$ is given by
$R_A$ in the present work, while it should be the confinement scale
$\sim 1\;$fm in reality.  We can specify $\bar{\mu}$ corresponding to
relevant $x$ through $Q_s(x)$ using Eq.~(\ref{eq:satu_MV}).

\begin{figure}
 \includegraphics[width=14cm]{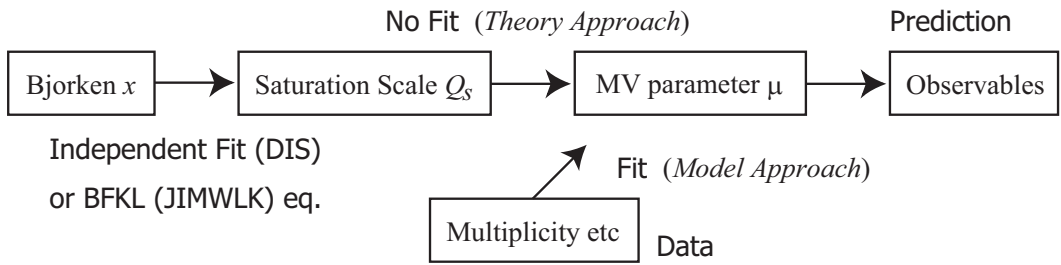}
 \caption{Schematic chart for the theory and model approaches.}
 \label{fig:chart}
\end{figure}

  We would call the above-mentioned strategy as the theory approach in
contrast to the model approach where $\bar{\mu}$ is phenomenologically
fixed by the total multiplicity.  Our point of view is summarized as a
chart in Fig.~\ref{fig:chart}.  This present work is a first-step
attempt to complete the full theory approach.

  Our finding that the longitudinal structure in infinitesimal extent
has a significant effect may well be important also for the
instability analyses in Ref.~\cite{Romatschke:2005pm}, as we already
mentioned in Introduction.  Because the instability takes place with
respect to the longitudinal fluctuations, it is naturally expected
that the proper treatment of longitudinal randomness should alter the
quantitative features discussed in Ref.~\cite{Romatschke:2005pm}.  It
is an intriguing question whether the instability would remain weak as
concluded in Ref.~\cite{Romatschke:2005pm} or not with longitudinal
randomness taken into account.  If not, the instability might be fast
and strong enough to account for the early thermalization.  We have to
emphasize that our $N_\eta$ is not the same as $L_\eta$ employed in
Ref.~\cite{Romatschke:2005pm} in which the authors introduced the
longitudinal \textit{coordinate} to solve the equation of motion in
three-dimensional space, but did not store the random sheets within
infinitesimal extent \textit{in the initial condition}.  The vital
difference lies in the fact that the number of the Monte-Carlo
integration variables $\bar{\rho}_n(\xt)$ ($n=1,\dots N_\eta$) becomes
greater with increasing $N_\eta$.  We conjecture that the Glasma
instability would become stronger with $N_\eta>1$, but the
quantitative analyses have to wait for at least $N_\eta$ times
expensive computations as compared to Ref.~\cite{Romatschke:2005pm},
which is the problem to be investigated in the future.

\acknowledgments
  The author thanks Teiji Kunihiro for useful conversations, Paul
Romatschke for encouraging and helpful comments, and Raju Venugopalan
for communications.

\end{document}